\documentclass[10pt,journal,twocolumn]{IEEEtran}
\usepackage{amssymb}
\usepackage{graphicx}
\usepackage{amsmath}
\usepackage{epsf}
\usepackage{mathrsfs}
\usepackage{cite}
\usepackage{dsfont}
\usepackage{float}
\usepackage{color}

 \usepackage[usenames,dvipsnames]{pstricks}
 \usepackage{epsfig}
 \usepackage{pst-grad} % For gradients
 \usepackage{pst-plot} % For axes
 \usepackage{pst-node}
 \usepackage{pst-3d}
 \usepackage{pstricks-add}

\newtheorem{theorem}{Theorem}

\newtheorem{definition}{Definition}

\newtheorem{remark}{Remark}

\def\antenna{%
\begin{pspicture}(1,1)
\pstriangle[gangle=-180.0](0,0.25)(0.6,0.25)
\psline(0,0)(0,-0.60)
\psline(-0.22,-0.60)(0.22,-0.60)
\end{pspicture}
}

\begin{document}
\title{Real Interference Alignment}
%\title{Signaling in Real Domains: Removing Barriers in Achieving the DOFs of Single Antenna Systems}

%\author{ Abolfazl~S.~Motahari,~\IEEEmembership{Student
%Member,~IEEE,}, ~Shahab Oveis Gharan,~\IEEEmembership{Student
%Member,~IEEE,},~ and~AmirE~K.~Khandani,~\IEEEmembership{Member,~IEEE}

\author{ Abolfazl~Seyed~Motahari$^{\dagger}$, ~Shahab Oveis-Gharan$^{\dagger}$,\\~Mohammad-Ali~Maddah-Ali$^{\dagger \dagger}$,~and~Amir~Keyvan~Khandani$^{\dagger}$
%\\\small Coding \& Signal Transmission Laboratory (www.cst.uwaterloo.ca)
\\\small $^{\dagger}$  Department of Electrical and Computer Engineering,  University of Waterloo, Waterloo, ON, Canada N2L3G1
\\ \{abolfazl,shahab,khandani\}@cst.uwaterloo.ca
\\ \small $^{\dagger\dagger}$ Department of Electrical Engineering and Computer Sciences, University of California-Berkeley, Berkeley, CA, USA
\\ maddah-a@eecs.berkeley.edu
}

%\author{
%\authorblockN{ Abolfazl S. Motahari and Amir~K.~Khandani}
%\authorblockA{Coding \& Signal Transmission Laboratory (www.cst.uwaterloo.ca)\\
%Dept. of Elec. and Comp. Eng., University of Waterloo \\
%Waterloo, ON, Canada N2L3G1 \\
%Email: (abolfazl,khandani)@cst.uwaterloo.ca}

%\and
%\authorblockN{Amir~K.~Khandani}
%\authorblockA{ECE Department \\
%University of Waterloo\\
%Waterloo, ON N2L3G1  \\
%Email: khandani@uwaterloo.ca}
%}

\maketitle

%\let\thefootnote\relax
%\footnotetext{Financial support provided by Nortel and the
%corresponding matching funds by the Natural Sciences and Engineering
%Research Council of Canada (NSERC), and Ontario Ministry of Research
%\& Innovation (ORF-RE) are gratefully acknowledged.}
%\let\thefootnote

\begin{abstract}
In this paper, we show that the total Degrees-Of-Freedoms (DOF) of the $K$-user Gaussian Interference Channel (GIC) can be achieved by incorporating a new alignment technique known as \emph{real interference alignment}. This technique compared to its ancestor \emph{vector interference alignment} performs on a single real line and exploits the properties of real numbers to provide optimal signaling. The real interference alignment relies on a new coding scheme in which several data streams having fractional multiplexing gains are sent by transmitters and interfering streams are aligned at receivers. The coding scheme is backed up by a recent result in the field of Diophantine approximation, which states that the convergence part of the Khintchine-Groshev theorem holds for points on non-degenerate manifolds.

\end{abstract}

\begin{keywords}
Interference channels, interference alignment, number theory, Diophantine approximation.
\end{keywords}

\section{Introduction}
\PARstart{A}{chieving} the optimum throughput of a system requires efficient interference management. Interference alignment is a type of interference management that exploits spatial Degrees-Of-Freedoms (DOF) available at transmitters and receivers. In \cite{maddahali2008com}, Maddah-Ali, Motahari, and Khandani introduced the concept of interference alignment and showed its capability in achieving the full Degrees-Of-Freedom (DOF) for certain classes of two-user $X$ channels.

Interference alignment in $n$-dimensional Euclidean spaces for $n\geq 2$ is studied by several researchers, c.f. \cite{maddahali2008com,jafar2008dfr,cadambe2008iaa}. In this method, at each receiver a subspace is dedicated to interference, then the signaling is designed such that all the interfering signals are squeezed in the interference subspace. Such an approach saves some dimensions for communicating desired signal, while keeping it completely free from the interference. Using this method, Cadambe and Jafar showed that, contrary to the popular belief, a $K$-user Gaussian interference channel with varying channel gains can achieve its total DOF, which is $\frac{K}{2}$. Later, in \cite{Bobak}, it is shown that the same result can be achieved using a simple approach based on a particular pairing of the channel matrices. The assumption of varying channel gains, particularly noting that all the gains should be known at the transmitters, is unrealistic, which limits the application of these important theoretical results in practice. This paper aims to remove this shortcoming by proving that the same result can be achieved by relying on a new alignment technique.

Some techniques are proposed for alignment in real domains. In~\cite{Bresler-Parekh-tse} interference alignment is applied in single antenna systems when all receivers are interference free but one. In  \cite{Etkin-Ordentlich} and \cite{abolfazl-shahab-amir}, the results from the field of Diophantine approximation in Number Theory are used to show that interference can be aligned using properties of rational and irrational numbers and their relations. However, in their examples there is no need for signaling design. The first example of interference alignment in one-dimensional spaces, which requires signaling design, is presented in \cite{abolfazl-real}. Using irrational numbers as transmit directions and applying Khintchine-Groshev theorem, \cite{abolfazl-real} shows the two-user $X$ channel achieves its total DOF. In this paper, we extend the result of \cite{abolfazl-real} and prove the following theorem
\begin{theorem}\label{main}
The total DOF of the $K$-user GIC with real and time invariant channel coefficients is $\frac{K}{2}$ for almost all channel realizations.
\end{theorem}

\begin{remark}
In \cite{Akbar-Abolfazl-Amir}, the total DOF of the $K$-user MIMO GIC is addressed. The application of real interference alignment to channels with complex coefficient gain is addressed in \cite{maddahalidof}. 
\end{remark}

%
%
%\textbf{Notation}: $\mathbb{R}$, $\mathbb{Q}$, $\mathbb{N}$
%represent the set of real, rational, and nonnegative integers,
%respectively. $(a,b)_{\mathbb{Z}}$ denotes the set of integers
%between $a$ and $b$.

\section{Real Interference Alignment}
The ingredients of real interference alignment technique are
\begin{enumerate}
\item Decoding based on Diophantine approximation.
\item Alignment based on structural encoding.
\item Achieving asymptotically perfect alignment based on partial interference alignment.
\end{enumerate}
We provide three examples to clarify the impact of the preceding items. We first look at a three-user multiple access channel modeled by $y=x_1+ax_2+bx_3+z.$ Let us assume that all three users communicate with the receiver using a
single data stream. The data streams are modulated by the constellation $\mathcal{U}=A(-Q,Q)_{\mathbb{Z}}$ where $A$ is a factor controlling the minimum distance of the received constellation. $(a,b)_{\mathbb{Z}}$ denotes the set of integers
between $a$ and $b$

The received constellation consists of points representable by $A(u_1+au_2+bu_3)$ where $u_i$s are integer. Let us choose two
distinct points $v_1=A(u_1+au_2+bu_3)$ and $v_2=A(u'_1+au'_2+bu'_3)$ in the received constellation. The distance between these two points
is $d=A|(u_1-u'_1)+a(u_2-u'_2)+b(u_3-u'_3)|$. The following theorem due to Groshev  can be used to lower bound the minimum distance of the received constellation.

\begin{theorem}[Khintchine-Groshev]\label{thm khintchine groshev}
The set of $m$-tuple real numbers satisfying
\begin{equation}\label{khintchine-distance}
 |p+\mathbf{v}\cdot \mathbf{q}|<\frac{1}{Q^{m+\epsilon}}
\end{equation}
has measure zero for $p\in \mathbb{Z}$, $\mathbf{q}\in\mathbb{Z}^m$, and $Q=\max\{|q_1|,\ldots,|q_m|\}$.
\end{theorem}

Put it differently, the theorem states that for almost all $\mathbf{v}\in\mathbb{R}^m$, there is a constant $\kappa$ which is only related to $\mathbf{v}$ such that $$|p+\mathbf{v}\cdot \mathbf{q}|>\frac{\kappa}{Q^{m+\epsilon}}$$ for all $p\in \mathbb{Z}$ and $\mathbf{q}\in\mathbb{Z}^m$.

Using the theorem, one can obtain $d_{\min}\approx \frac{A}{Q^{2}}$ where $d_{\min}$ is the minimum distance in the received constellation. It can be shown that if $d_{\min}\approx 1$ then the additive gaussian noise can be removed from the received signal using an appropriate coding scheme \cite{khodam}. This condition can be enforced by $A\approx Q^2$. In a noise-free environment, the receiver can decode  the three messages if there is a one-to-one
map from the received signal to the transmit constellation. This condition can be satisfied if $a$ and $b$ are rationally independent which in fact holds almost surely. Therefore the receiver can decode all three messages almost surely.

To calculate User $i$'s rate $R_i=\log(2Q-1)$ in terms of $P$, we need to find a relation between $Q$ and $P$. Due to the power constraints, we have $P=A^2Q^2$. We showed that $A\approx Q^2$. Therefore, $P\approx Q^6$. Hence, we have
\begin{equation}
r_i=\lim_{P\rightarrow\infty}\frac{R_i}{0.5\log P}=\frac{1}{3},
\end{equation}
where $r_i$ is the achievable DOF for User $i$. 

In the preceding example, we implicitly assumed that the pair $(a,b)$ can take any value in $\mathbb{R}^2$. Otherwise, we were not able to apply the Khintchine-Groshev theorem. Let us assume that $a$ and $b$ have a relation. For instance, $b$ is a function of $a$, say $b=a^2$. In this case, the pair $(a,b)$ lies on a one dimensional manifold in $\mathbb{R}^2$, see Figure \ref{fig bad-events}. Since the manifold itself has measure zero, Khintchine-Groshev theorem can
not be applied directly. For such cases, however, there is an extension to Khintchine-Groshev theorem , see \cite{groshev-berink} and \cite{groshev-Beresnevich}, which states that the same lower bound on the minimum distance can be applied when coefficients lie on a non-degenerate manifold and, in fact, the measure of points not satisfying the theorem is zero. In can be shown that if all $v_i$'s in (\ref{khintchine-distance}) are monomials with variables from the set $\mathbf{g}=\{g_1,g_2,\ldots,g_n\}$, then $\mathbf{v}$ lines on a non-degenerate manifold. As a special case when set $\mathbf{v}$
has only one member, i.e. $\mathbf{v}=\{1,g,g^2,g^3,\ldots\}$.

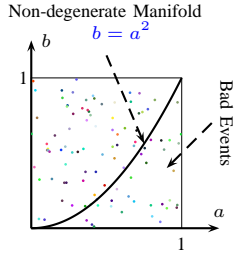
\begin{figure}
\centering
\scalebox{1}
{
\begin{pspicture}(0,0)(3,3)
%\psgrid
 \rput(-.1,2){\scriptsize 1}
 \rput(2,-.2){\scriptsize 1}
 \psline{->}(0,2.5)
 \psline{->}(2.5,0)
 \rput(2.5,.2){\scriptsize $a$}
 \rput(.2,2.5){\scriptsize $b$}
 \psRandom[dotsize=1pt,color,fillstyle=solid,randomPoints=100](0,0)(2,2){\psframe[linewidth=0](0,0)(2,2)}
 \psline[linestyle=dashed]{->}(2.4,1.4)(1.8,.8)
 \rput{-90}(2.6,1.4){\scriptsize Bad Events}
 \rput(1.2,2.75){\parbox{3cm}{\scriptsize  Non-degenerate Manifold \\ \centerline{$\textcolor[rgb]{0.00,0.00,1.00}{b=a^2}$}}}
 \psline[linestyle=dashed]{->}(1,2.3)(1.5,1.125)
 \psdot[dotsize=1pt](1.4,1.3)
 \psplot[plotpoints=100]{0}{2}{x x mul 2 div}
\end{pspicture}
}
\caption{There are infinitely many points on the $a-b$ plane with
measure zero that are not addressed by the Khintchine-Groshev
theorem (these are called bad events). The curve $b=a^2$ is a
non-degenerate manifold and by the extension of Khintchine-Groshev
theorem, the measure of bad events is zero on the curve. }\label{fig
bad-events}
\end{figure}

In real interference alignment we say two data streams are aligned at a
receiver if they arrive at the receiver with the same received
direction (coefficient). To show this in fact reduces the total dimension of the received interference, we consider  the two-user $X$ channel. In the two-user $X$ channel, each transmitter has independent messages to both receivers, see Figure \ref{fig x channel}. Hence, each transmitter has two data streams and they need to be
transmitted such that they can be separated in their corresponding
receivers. In \cite{abolfazl-real}, the following signaling is
proposed for the channel:
\definecolor{vcolor}{rgb}{0.00,1.00,1.00}
\definecolor{ucolor}{rgb}{1.00,0.50,1.00}
\begin{IEEEeqnarray*}{rl}
x_1&=h_{22} \textcolor{red}{u_1}+h_{12} \textcolor{blue}{v_1},\\
x_2&=h_{21} \textcolor{red}{u_2}+h_{11} \textcolor{blue}{v_2},
\end{IEEEeqnarray*}
where $u_1,u_2$ and $v_1,v_2$ are data streams intended for the
first and second receivers, respectively. All data streams are
transmitted using the constellation
$\mathcal{U}=A(-Q,Q)_{\mathbb{Z}}$, where $Q$ is an integer and $A$
is the factor controlling the minimum distance of the received
constellation.

Using the above signaling, the received signal can be written as:
\begin{IEEEeqnarray*}{rl}
y_1&=(h_{11}h_{22}) \textcolor{red}{u_1}+(h_{21}h_{12})\textcolor{red}{u_2} +(h_{11}h_{12})(\textcolor{blue}{v_1}+\textcolor{blue}{v_2})+z_1,\\
y_2&=(h_{21}h_{12}) \textcolor{blue}{v_1}+(h_{11}h_{22})\textcolor{blue}{v_2} +(h_{21}h_{22})(\textcolor{red}{u_1}+\textcolor{red}{u_2})+z_2.
\end{IEEEeqnarray*}
The received signals are linear combinations of three terms in which
two of them are the intended data streams and one is the sum of
interfering signals, see Figure \ref{fig x channel}. Let us focus on
the first receiver. $y_1$ resembles the received signal of a
multiple access channel with three users. However, there is an
important difference between them. In the two-user $X$ channel the
term corresponding to the interfering signals, i.e. $u_3=v_1+v_2$,
is a sum of two data streams. However, we claim that this difference
does not change considerably the minimum distance of the received
constellation, i.e. $d_{\min}$. Recall that Khintchine-Groshev
theorem is used to bound $d_{\min}$. The bound is a function of the
maximum value that the integers can take. The maximum value of $u_3$
is $2AQ$, which is different from a single data stream by a factor
of two. Since this change only affects the constant term of
Khintchine-Groshev theorem, we have $d_{\min}\approx \frac{A}{Q^2}$
and the receiver can decode all data streams if each of them have a
multiplexing gain of $\frac{1}{3}$. Therefore, the multiplexing gain
of $\frac{4}{3}$ is achievable in total, which meets the upper
bound.

\def\antenna{\psline(0.05,0)(.35,0)
\psline(.2,0)(.2,0.5)
\pspolygon(.2,0.5)(.4,.7)(0,.7)}

\begin{figure}
\centering
\begin{pspicture}(-2.4,-.1)(5,2.8)
%\psgrid
\rput(-.3,1.9){\antenna}
\rput(1.8,1.9){\antenna}
\rput(1.8,-0.1){\antenna}
\rput(-.3,-0.1){\antenna}
\rput(-1,2.3){\psline{->}(.7,0) \rput(.3,.2){\scriptsize $x_1$}}
\rput(-1,0.3){\psline{->}(.7,0)\rput(.3,.2){\scriptsize $x_2$}}
\rput{180}(3,2.3){\psline{<-}(.7,0)\rput{180}(.3,-.2){\scriptsize $y_1$}}
\rput{180}(3,.3){\psline{<-}(.7,0)\rput{180}(.3,-.2){\scriptsize $y_2$}}
\cnode(0.2,2.3){0}{T1}
\cnode(0.2,0.3){0}{T2}
\cnode(1.8,2.3){0}{R1}
\cnode(1.8,0.3){0}{R2}
\ncline{->}{T1}{R1}
\ncput*[nrot=:U]{\tiny{$h_{11}$}}
\ncline{->}{T2}{R1}
\ncput*[npos=.7,nrot=:U]{\tiny{$h_{12}$}}
\ncline{->}{T1}{R2}
\ncput*[npos=.7,nrot=:U]{\tiny{$h_{21}$}}
\ncline{->}{T2}{R2}
\ncput*[nrot=:U]{\tiny{$h_{22}$}}

\def\mypoint #1#2#3{\pscircle[linecolor=#3,fillcolor=#3,fillstyle=solid](#1){.2}
\rput(#1){\scriptsize $\textcolor{black}{#2}$}}

\def\mymultipoint #1#2{\rput(#1){
\psframe(-.25,-.25)(.25,1.25)
\mypoint{0,0}{1}{#2}
\mypoint{0,.5}{2}{#2}
}}

\mypoint{-1.4,.3}{v_2}{blue}
\mypoint{-2.1,.3}{u_2}{red}
\mypoint{-1.4,2.3}{v_1}{blue}
\mypoint{-2.1,2.3}{u_1}{red}

\mypoint{4.8,.3}{v_2}{blue}
\mypoint{4.8,2.3}{u_2}{red}
\mypoint{4.1,.3}{v_1}{blue}
\mypoint{4.1,2.3}{u_1}{red}

\rput(4.8,1.8){\scriptsize $\frac{1}{3}$}
\rput(4.1,1.8){\scriptsize $\frac{1}{3}$}
\rput(3.4,1.8){\scriptsize $\frac{1}{3}$}

\rput(4.8,-.2){\scriptsize $\frac{1}{3}$}
\rput(4.1,-.2){\scriptsize $\frac{1}{3}$}
\rput(3.4,-.2){\scriptsize $\frac{1}{3}$}

\pscircle[fillcolor=black,fillstyle=solid](3.4,2.3){.2}
\rput(3.4,2.6){\scriptsize}

\rput(3.4,2.3){
\psframe(-.25,-.25)(.25,.75)
\mypoint{0,0}{v_1}{blue}
\mypoint{0,.5}{v_2}{blue}
}

\rput(3.4,.3){
\psframe(-.25,-.25)(.25,.75)
\mypoint{0,0}{u_1}{red}
\mypoint{0,.5}{u_2}{red}
}

\end{pspicture}
\caption{The two-user $X$ channel. }\label{fig x channel}
\end{figure}
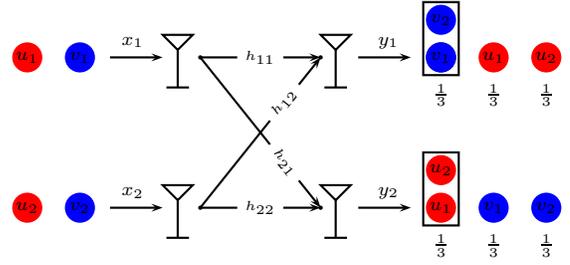

The signaling proposed for the two-user $X$ channel can be
interpreted as follows. The received signal at each receiver is a
real number, which is a one-dimensional component. One can embed
three rational dimensions, each of which has dimension $\frac{1}{3}$
in this one dimensional space, see Figure \ref{fig x channel}. One
of these dimensions is associated with interference and the other
two with intended signals. Therefore, $\frac{4}{3}$ out of two
dimensions available at both receivers are used for data, which in
turn gives us the total DOF of the channel.

In the two user $X$ channel, we have observed that
interfering signals from two different sources can be easily aligned
at a single receiver. Moreover, two interfering streams are received
with the same direction occupying only $\frac{1}{3}$ of the
available dimensions of the receivers. This is in fact the best
efficiency that one can hope for in reducing the number of waste
dimensions. This perfect alignment is not possible in general. In the following example, we  show that partial alignment can be used instead to provide the same performance asymptotically.

\begin{figure}
\centering
\begin{pspicture}(-1.8,-2.2)(3.8,2.8)
%\psgrid
\rput(-.3,1.9){\antenna}
\rput(1.8,.9){\antenna}
\rput(1.8,-1.1){\antenna}
\rput(-.3,-0.1){\antenna}
\rput(-.3,-2.1){\antenna}

\rput(-1,2.3){\psline{->}(.7,0) \rput(.3,.2){\scriptsize $x_1$}}
\rput(-1,0.3){\psline{->}(.7,0)\rput(.3,.2){\scriptsize $x_2$}}
\rput(-1,-1.7){\psline{->}(.7,0)\rput(.3,.2){\scriptsize $x_3$}}
\rput{180}(3,1.3){\psline{<-}(.7,0)\rput{180}(.3,-.2){\scriptsize $y_1$}}
\rput{180}(3,-.7){\psline{<-}(.7,0)\rput{180}(.3,-.2){\scriptsize $y_2$}}

\cnode(0.2,2.3){0}{T1}
\cnode(0.2,0.3){0}{T2}
\cnode(0.2,-1.7){0}{T3}
\cnode(1.8,1.3){0}{R1}
\cnode(1.8,-0.7){0}{R2}

\ncline{->}{T1}{R1}
\ncput*[npos=.7]{\tiny{$1$}}
\ncline{->}{T2}{R1}
\ncput*[npos=.7]{\tiny{$1$}}
\ncline{->}{T3}{R1}
\ncput*[npos=.8]{\tiny{$1$}}
\ncline{->}{T1}{R2}
\ncput*[npos=.8]{\tiny{$a$}}
\ncline{->}{T2}{R2}
\ncput*[npos=.7]{\tiny{$b$}}
\ncline{->}{T3}{R2}
\ncput*[npos=.7]{\tiny{$c$}}

\def\mypoint #1#2#3{\pscircle[linecolor=#3,fillcolor=#3,fillstyle=solid](#1){.2}
\rput(#1){\scriptsize $\textcolor{black}{#2}$}}

\def\mymultipoint #1#2{\rput(#1){
\psframe(-.25,-.25)(.25,1.25)
\mypoint{0,0}{1}{#2}
\mypoint{0,.5}{2}{#2}
\mypoint{0,1}{3}{#2}
}}

\pscircle[fillcolor=red,fillstyle=solid](-1.4,.3){.2}
\rput(-1.4,.7){\scriptsize $\textcolor{red}{\mathcal{T}}$}
\pscircle[fillcolor=blue,fillstyle=solid](-1.4,2.3){.2}
\rput(-1.4,2.7){\scriptsize $\textcolor{blue}{\mathcal{T}}$}
\pscircle[fillcolor=green,fillstyle=solid](-1.4,-1.7){.2}
\rput(-1.4,-1.3){\scriptsize $\textcolor{green}{\mathcal{T}}$}

\rput(3.4,-.7){
\psframe(-.25,-.25)(.25,1.25)
\mypoint{0,0}{a\mathcal{T}}{blue}
\mypoint{0,.5}{b\mathcal{T}}{red}
\mypoint{0,1}{c\mathcal{T}}{green}
}

\rput(3.4,1.3){
\psframe(-.25,-.25)(.25,1.25)
\mypoint{0,0}{\mathcal{T}}{blue}
\mypoint{0,.5}{\mathcal{T}}{red}
\mypoint{0,1}{\mathcal{T}}{green}
}

\end{pspicture}
\caption{Three transmitters wish to align their signals at two
receivers.}\label{fig alignment at two}
\end{figure}
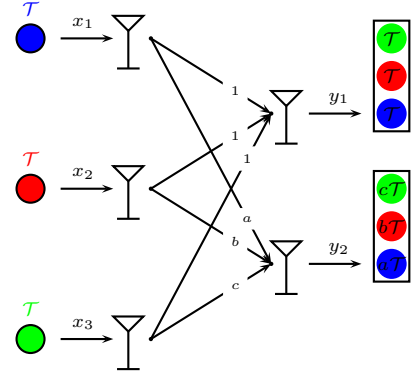

Let us consider a communication scenario in which three transmitters
try to align their signals at two different receivers. The channel
is depicted in Figure \ref{fig alignment at two}. In order to shed
light on the alignment part of the signaling, the intended receivers
are removed from the picture.

Alignment can be done at the first receiver by sending a single data
stream with direction $1$ from each of the transmitters; whereas
alignment at the second receiver requires $bc$, $ac$, and $ab$ as
chosen transmit directions for first, second, and third
transmitters, respectively. In general, it is not possible to
simultaneously align three single data streams at two different
receivers. Therefore perfect alignment is not feasible by
transmitting single data streams from each transmitter.

The solution to this problem is partial alignment, which is first
introduced in \cite{cadambe2008iaa}. In this technique, instead of
sending just one data stream, several data streams are transmitted
from each transmitter. The idea is to choose the transmit directions
based on channel coefficients in such a way that the number of
received directions is minimum. For the sake of simplicity, we choose
the same directions at all transmitters. Let $\mathcal{T}$ denote
the set of transmit directions. A direction $T\in\mathcal{T}$ is
chosen as a transmit direction if it can be represented as
\begin{equation}
T=a^{s_{1}}b^{s_{2}}c^{s_{3}},
\end{equation}
where $0\leq s_{i}\leq n-1$ for all $i\in\{1,2,3\}$. In this way,
the total number of transmit directions is $L_1=n^3$.

To compute the efficiency of the alignment, one needs to find the
set of received directions in the first and second receivers, which
are denoted by $\mathcal{T}_1$ and $\mathcal{T}_2$, respectively.
Since all transmit directions arrive at the first receiver intact,
$\mathcal{T}_1=\mathcal{T}$.

To compute the set of received directions at the second receiver, we
look at the received directions due to the first, second, and third
transmitters separately. Since all of them are multiplied by $a$,
the received directions due to the first transmitter are of the form
$a^{s_{l}+1}b^{s_{2}}c^{s_{3}}$, where $0\leq s_{i}\leq n-1$ for all
$i\in\{1,2,3\}$. Similarly, $a^{s_{l}}b^{s_{2}+1}c^{s_{3}}$ and
$a^{s_{l}}b^{s_{2}}c^{s_{3}+1}$ are the types of received directions
due to the second and third transmitter, respectively. Taking the
union of all these directions, one can compute $\mathcal{T}_2$.
However, we can easily see that the set of directions formed by
$a^{s_{l}}b^{s_{2}}c^{s_{3}}$, where  $0\leq s_{i}\leq n$ for all
$i\in\{1,2,3\}$ includes $\mathcal{T}_2$ and can be used as an upper
bound on the number of received directions. This set has $(n+1)^3$,
which is an upper bound for $L_2$. Hence, we conclude that
$$\eta=\frac{L_1}{L_2}>\left(\frac{n}{n+1}\right)^3.$$ Since $n$ is
an arbitrary integer, any alignment efficiency close to 1 is
possible. Hence, the  partial alignment approaches the perfect
alignment.

For the multiple transmitter and receiver, the above approach can be
easily extended. In fact, it can be shown that asymptotically perfect alignment
is possible for any finite number of transmitters and receivers.

In the following section, we need the following theorem which in fact summarizes the conditions needed to achieve
the multiplexing gain of $\frac{1}{m}$ per data stream in a system. For the proof please see \cite{khodam}.
\begin{theorem}\label{basic1}
Consider there are $K$ transmitters and $K'$ receivers in a system
parameterized by the channel coefficient vector $\mathbf{h}$.
Transmitter $i$ sends $M$ data stream along directions
$\mathcal{T}_i=\{T_{i0},T_{i2},\ldots,T_{i(M-1)}\}$ for all $i\in
\{1,2,\ldots,K\}$. The data streams intended for the $j$'th receiver
arrive at $L_j$ directions, which are
$\mathcal{T}_j=\{\bar{T}_{j0},\bar{T}_{j2},\ldots,\bar{T}_{j(L_j-1)}\}$.
Moreover, the interference part of the received signal at the $j$'th
receiver has $L'_j$ effective data streams with received directions
$\mathcal{T}'_j=\{\bar{T}'_{j0},\bar{T}'_{j2},\ldots,\bar{T}'_{j(L'_j-1)}\}$
for all $j\in \{1,2,\ldots,K'\}$. Let the following conditions for
all $j\in \{1,2,\ldots,K'\}$ hold:
\begin{description}
  \item[C1] Components of $\mathcal{T}_i$ are distinct member of $\mathcal{G}(\mathbf{h})$ and linearly independent over the field of rational numbers.
  \item[C2] Components of $\mathcal{T}_{i}$ and $\mathcal{T}'_{i}$ are all distinct.
  \item[C3] One of the elements of either $\mathcal{T}_{i}$ or $\mathcal{T}'_{i}$ is 1.
\end{description}
Then, by encoding each data stream using the constellation $\mathcal{U}=(-Q,Q)_{\mathbb{Z}}$ where $Q=\gamma
P^{\frac{1-\epsilon}{2(m+\epsilon)}}$ and $\gamma$ is a constant, the following DOF is achievable for almost all realizations of the system:
\begin{equation}
r_{\text{sum}}=\frac{L_1+L_2+\cdots+L_{K'}}{m},
\end{equation}
where $m$ is the maximum received directions among all receivers, i.e., $m=\max_i L_i+L'_i$.
\end{theorem}

\begin{remark}
If C3 does not hold, then by adding a virtual data stream in the
direction 1 at the receiver, one can conclude that $\frac{1}{m+1}$
is achievable for all data streams.
\end{remark}

\section{Proof of the Main Theorem}

\subsection{System Model}\label{sec system model}

The $K$-user GIC models a network in which $K$ transmitter-receiver
pairs (users) sharing a common bandwidth wish to have reliable
communication at their maximum rates. The channel's input-output
relation can be stated as follows, %see Figure \ref{k-user IC},
\begin{IEEEeqnarray}{rl}\label{k-user model}
 y_1 &=h_{11}x_1+h_{12}x_2+\ldots +h_{1K}x_K+z_1,\nonumber\\
 y_2 &=h_{21}x_1+h_{22}x_2+\ldots +h_{2K}x_K+z_2,\nonumber\\
\vdots\ &=\quad  \vdots \quad\qquad \vdots \quad\qquad\ddots\qquad\vdots \\
y_K &=h_{K1}x_1+h_{K2}x_2+\ldots +h_{KK}x_K+z_K,\nonumber
\end{IEEEeqnarray}
where $x_i$ and $y_i$ are input and output symbols of User $i$ for
$i\in\{1,2,\ldots,K\}$, respectively. $z_i$ is Additive White Gaussian Noise (AWGN) with unit
variance for $i\in\{1,2,\ldots,K\}$. Transmitters are subject to the
power constraint $P$. $h_{ji}$ represents the channel gain between
Transmitter $i$ and Receiver $j$. It is assumed that all channel
gains are real and time invariant. The set of all channel gains is
denoted by $\mathbf{h}$, i.e.,
$\mathbf{h}=\{h_{11},\ldots,h_{1K},h_{21},\ldots,h_{2K},\ldots,h_{K1},\ldots,h_{KK}\}$.
Since the noise variances are normalized, the Signal to Noise Ratio
(SNR) is equivalent to the input power $P$. Hence, we use them
interchangeably throughout this paper.

In this paper, we are primarily interested in characterizing the
total DOF of the $K$-user GIC. Let $\mathcal{C}$ denote the capacity
region of this channel. The DOF region  associated with the channel
is in fact the shape of $\mathcal{C}$ in high SNR regimes scaled by
$\log \text{SNR}$. Let us denote the DOF region by $\mathcal{R}$.
All extreme points of $\mathcal{R}$ can be identified by solving the
following optimization problem:
\begin{equation}
r_{\boldsymbol{\lambda}}=\lim_{\text{SNR}\rightarrow\infty}\max_{\mathbf{R}\in
\mathcal{C}}\frac{\boldsymbol{\lambda}^t \mathbf{R}}{0.5\log
\text{SNR}}.
\end{equation}
The total DOF refers to the case where
$\boldsymbol{\lambda}=\{1,1,\ldots,1\}$, i.e., the sum-rate is
concerned. Throughout this paper, $r_{\text{sum}}$ denotes the total
DOF of the system.

An upper bound on the DOF of this channel is obtained in
\cite{cadambe2008iaa}. The upper bound states that the total DOF of
the channel is less than $\frac{K}{2}$, which means each user can at
most enjoy one half of its maximum DOF.

\subsection{Three-user Gaussian Interference Channel:  $\text{DOF}=\frac{3}{2}$ is Achievable}\label{sec three-user}
In this section, we consider the three-user GIC and  explain in
detail that, by an appropriate selection of transmit directions, the
DOF of $\frac{3}{2}$ is achievable for almost all cases. We will
explain in more detail that by an appropriate selection of transmit
directions this DOF can be achieved.

In \cite{abolfazl-real}, we defined the standard model of the three-user GIC. The
definition is as follows:
\begin{definition}
The three user interference channel is called standard if it can be represented as
\begin{IEEEeqnarray}{rl}\label{alaki7}
y_1&=G_1x_1+x_2+x_3+z_1\nonumber\\
y_2&=G_2x_2+x_1+x_3+z_2\\
y_3&=G_3x_3+x_1+G_0x_2+z_3.\nonumber
\end{IEEEeqnarray}
where $x_i$ for User $i$ is subject to the power constraint $P$.
$z_i$ at Receiver $i$ is AWGN with unit variance.
\end{definition}
In \cite{abolfazl-real}, it is also proved that every three-user GIC
has an equivalent standard channel as far as the DOF is concerned.

As mentioned in the previous section, transmit directions are
monomials with variables from channel coefficients. For the three
user case, we only use $G_0$ as the generator of transmit
directions. Therefore, transmit directions are selected from the set
$\mathcal{G}(G_0)$, which is a subset of
$\mathcal{G}(G_0,G_1,G_2,G_3)$. Clearly,
$\mathcal{G}(G_0)=\{1,G_0,G_0^2,G_0^3,\cdots\}$.

We only consider the case where $G_0$ is transcendental. In fact, the measure of being
algebraic is zero. If $G_0$ is transcendental then all members of $\mathcal{G}(G_0)$
are linearly independent over the field of rational numbers. Hence,
we are not limited to any subset of $\mathcal{G}(G_0)$, as far as
the independence of transmit directions is concerned. We will show
that $\frac{3n+1}{2n+1}$ is an achievable DOF for any
$n\in\mathbb{N}$. To this end, we propose a design that is not
symmetrical.

Transmitter 1 uses the set of directions
$\mathcal{T}_1=\{1,G_0,G_0^2,\ldots,G_0^{n}\}$ to transmit $L_1=n+1$
to its corresponding receiver. Clearly $\mathcal{T}_1$ satisfies C1.
The transmit signal from User 1 can be written as
$$x_1=A\sum_{j=0}^n G_0^j u_{1j}.$$ Transmitters 2 and 3 transmit in
$L_2=L_3=n$ directions using
$\mathcal{T}_2=\mathcal{T}_3=\{1,G_0,G_0^2,\ldots,G_0^{n-1}\}$.
Clearly both $\mathcal{T}_2$ and $\mathcal{T}_3$ satisfy C1. The
transmit signals can be expressed as $$x_2=A\sum_{j=0}^{n-1} G_0^j
u_{2j}$$ and $$x_3=A\sum_{j=0}^{n-1} G_0^j u_{3j}.$$

The received signal at Receiver 1 can be expressed as:
\begin{equation}
 y_1=A\left(\sum_{j=0}^n G_1G_0^j u_{1j}+\sum_{j=0}^{n-1} G_0^j u'_{1j}\right)+z_1,
\end{equation}
where $u'_{1j}=u_{2j}+u_{3j}$. In fact, transmit signals from Users 2 and 3 are aligned at Receiver 1. This is
due to the fact that out of $2n$ possible received directions, only $n$ directions are effective, i.e.,
$L'_1=n$. One can also confirm that C2 and C3 are held at Receiver 1.

The received signal at Receiver 2 can be expressed as:
\begin{equation}
 y_2=A\left(\sum_{j=0}^{n-1} G_2G_0^j u_{2j}+\sum_{j=0}^{n} G_0^j u'_{2j}\right)+z_2,
\end{equation}
where $u'_{2j}=u_{1j}+u_{3j}$ for all $j\in\{0,1,\ldots,n-1\}$ and
$u'_{2n}=u_{1n}$. At Receiver 2, transmitted signals from Users 1
and 3 are aligned and the number of effective received directions is
$L'_2=n+1$. Moreover, it can be easily seen that C2 and C3 hold at
Receiver 2.

The received signal at Receiver 3 can be expressed as:
\begin{equation}
 y_3=A\left(\sum_{j=0}^{n-1} G_3G_0^j u_{3j}+\sum_{j=0}^{n} G_0^j u'_{3j}\right)+z_3,
\end{equation}
where $u'_{3j}=u_{1j}+u_{2j}$ for all $j\in\{1,2,\ldots,n\}$ and
$u'_{30}=u_{10}$. At Receiver 3, transmitted signals from Users 1
and 2 are aligned and the number of effective received directions is
$L'_2=n+1$. Clearly, C2 and C3 hold for Receiver 3.

Since C1, C2, and C3 hold at all users, we only need to obtain the
number of maximum received directions at all receivers. To this end,
we observe that $$m=\max\{L_1+L'_1,L_2+L'_2,L_3+L'_3\}=2n+1$$.
Therefore, an application of Theorem \ref{basic1} reveals that the
following DOF is achievable.
\begin{IEEEeqnarray}{rl}
r_{\text{sum}} & =\frac{L_1+L_2+L_3}{m}\nonumber\\
& =\frac{3n+1}{2n+1}.
\end{IEEEeqnarray}
Since $n$ is an arbitrary integer, one can conclude that
$\frac{3}{2}$ is achievable for the three-user GIC almost surely.

\subsection{$K$-user Gaussian Interference Channel: $\text{DOF}=\frac{K}{2}$ is Achievable}\label{sec k-user}
To prove the main result of the paper, we start with selecting the transmit directions for User $i$. A
direction $T\in\mathcal{G}(\mathbf{h})$ is chosen as the transmit
direction for User $i$ if it can be represented as
\begin{equation}\label{k-user transmit directions}
T=\prod_{j=1}^{K}\prod_{l=1}^{K}  h_{jl}^{s_{jl}},
\end{equation}
where $s_{jl}$'s are integers satisfying
\begin{equation}\nonumber
\begin{cases}
s_{jj}=0 & \forall ~ j\in\{1,2,\ldots,K\}\\
0\leq s_{ji}\leq n-1 &\forall ~ j\in\{1,2,\ldots,K\} ~\& ~j\neq i\\
0\leq s_{jl}\leq n & ~ \text{Otherwise.}
\end{cases}
\end{equation}
The set of all transmit directions is denoted by $\mathcal{T}_i$. It is easy to show that the cardinality of this set is \begin{equation}
L_i=n^{K-1}(n+1)^{(K-1)^2}.
\end{equation}
Clearly, $\mathcal{T}_i$ satisfies C1 for all
$i\in\{1,2,\ldots,K\}$.

To compute $L'_i$ (the number of independent received directions due
to interference), we investigate the effect of Transmitter $k$ on
Receiver $i$. Let us first define $\mathcal{T}_r$ as the set of
directions represented by (\ref{k-user transmit directions}) and
satisfying
\begin{equation}\label{k-user condition1}
\begin{cases}
s_{jj}=0 & \forall ~ j\in\{1,2,\ldots,K\}\\
0\leq s_{jl}\leq n & ~ \text{Otherwise.}
\end{cases}
\end{equation}
We claim that $\mathcal{T}_{ik}$, the set of received directions at
Receiver $i$ due to Transmitter $k$, is a subset of $\mathcal{T}_r$.
In fact, all transmit directions of Transmitter $k$ arrive at
Receiver $i$ multiplied by $h_{ik}$. Based on the selection of
transmit directions, however, the maximum power of $h_{ik}$ in all
members of $\mathcal{T}_{ik}$ is $n-1$. Therefore, none of the
received directions violates the condition of (\ref{k-user
condition1}) and this proves the claim.

Since $\mathcal{T}_r$ is not related to User $k$, one can conclude
that $\mathcal{T}_{ik}\subseteq \mathcal{T}_r$ for all
$k\in\{1,2,\ldots,K\}$ and $k\neq i$. Hence, we deduce that all
interfering users are aligned in the directions of $\mathcal{T}_r$.
Now, $L'_i$ can be obtained by counting the members of
$\mathcal{T}_r$. It is easy to show that
\begin{equation}
L'_i=(n+1)^{K(K-1)}.
\end{equation}

The received directions at Receiver $i$ are members of
$h_{ii}\mathcal{T}_i$ and $\mathcal{T}_r$. Since $h_{ii}$ does not
appear in members of $\mathcal{T}_r$, the members of
$h_{ii}\mathcal{T}_i$ and $\mathcal{T}_r$ are distinct. Therefore,
C2 holds at Receiver $i$. Since all the received directions are
irrationals, C3 does not  hold at Receiver $i$.

Since $C_1$ and $C_2$ hold for all users, we can apply Theorem \ref{basic1} to obtain the DOF of the channel. We have
\begin{IEEEeqnarray}{rl}
r_{\text{sum}}&=\frac{L_1+L_2+\ldots+L_K}{m+1}\nonumber\\
&=\frac{Kn^{K-1}(n+1)^{(K-1)^2}}{m+1}
\end{IEEEeqnarray}
where $m$ is
\begin{IEEEeqnarray}{rl}
 m &=\max_i L_i+L'_i\nonumber\\
   &=n^{K-1}(n+1)^{(K-1)^2}+(n+1)^{K(K-1)}.
\end{IEEEeqnarray}
Combining the two equations, we obtain
\begin{equation}
r_{\text{sum}}=\frac{K}{1+(\frac{n+1}{n})^{K-1}+\frac{1}{n^{K-1}(n+1)^{(K-1)^2}}}.
\end{equation}
Since $n$ can be arbitrary large, we conclude that $\frac{K}{2}$ is achievable for the $K$-user GIC.

%\bibliographystyle{IEEEtran}
%\bibliography{references}
% Generated by IEEEtran.bst, version: 1.12 (2007/01/11)

\end{document}